\documentclass[10pt,conference]{IEEEtran}
\usepackage[absolute,overlay]{textpos} % For fixed positioning of the copyright notice
\usepackage{ragged2e}

\usepackage{graphicx} % Required for inserting images
\usepackage{physics}
\usepackage{tikz}
\usetikzlibrary{calc}
\usepackage{xcolor}
\definecolor{darkred}{rgb}{0.5, 0, 0}
\usepackage{hyperref}
\usepackage{subcaption}
\usepackage{amsmath}
\usepackage{amssymb}
\usepackage{bbm}
\usepackage{cite}
\newtheorem{Protocol}{Protocol}
\usepackage[onehalfspacing]{setspace}
\definecolor{greenish}{HTML}{228833}

\title{How to share Multipartite Entanglement in a Real-World Linear Network Connecting Two Metropoles}
\date{\today}

\begin{document}

% Authors and Affiliations
\author{
    \IEEEauthorblockN{Janka Memmen, Anna Pappa}  % List of author names
    \IEEEauthorblockA{%
        \textit{Electrical Engineering and Computer Science Department}\\
        Technische Universit{\"a}t Berlin, 10587 Berlin, Germany\\
        %Email: janka.memmen@tu-berlin.de
    }}

\maketitle
\begin{textblock*}{18.2cm}(1.7cm,26.5cm)  % width, position (X, Y)
\footnotesize
\justifying
\noindent
\copyright~2025 IEEE. Personal use of this material is permitted. Permission from IEEE must be obtained for all other uses, in any current or future media, including
reprinting/republishing this material for advertising or promotional purposes, creating new collective works, for resale or redistribution to servers or lists, or
reuse of any copyrighted component of this work in other works.
\end{textblock*}

\noindent \begin{abstract}
    The development of large-scale quantum communication networks necessitates the efficient distribution of quantum states to enable advanced cryptographic applications and distributed tasks. Multipartite entanglement is a key resource in many of these proposals, yet its generation is experimentally challenging, especially in noisy and lossy networks. While a substantial body of work focuses on the distribution of multi-partite entanglement in star-like topologies, practical implementations often rely on linear network structures constrained by existing infrastructure. In this work, we investigate the generation of high-fidelity multipartite entangled states in a realistic quantum network, leveraging the existing infrastructure of the Q-net-Q project—a real-world long-distance link connecting Berlin and Frankfurt via seven trusted relay nodes. Given that only bipartite entanglement sources are available in our setting and that the network is highly lossy, we explore the role of quantum memories in enhancing multi-partite entanglement distribution and identify key performance requirements. Furthermore, we analyze the feasibility and performance of cryptographic primitives—including (Anonymous) Conference Key Agreement and Quantum Secret Sharing—highlighting the scenarios where the use of multipartite entanglement yields clear advantages.
\end{abstract}
\section{Introduction}
In recent years, large-scale quantum communication networks involving multiple participants have gained significant attention \cite{Kimble:2008, Wehner:2018, vidick:2023}. A natural resource for these networks is multi-partite entanglement, which enables advanced cryptographic protocols and distributed quantum computing while simultaneously offering advantages in terms of distribution times and memory usage for certain network topologies.
The distribution of multi-partite entanglement in star-like network topologies has been extensively studied \cite{Epping:2016hc, Epping_2017, Avis:2023jy, memmen2023}; however many real-world networks do not naturally provide this structure. Instead, practical quantum networks are often linear and additionally constrained by the existing infrastructure and available resources.

\begin{figure*}[!ht]
    \centering
    \begin{tikzpicture}[scale=1.4]

        % Define styles
        \tikzstyle{endnode} = [rectangle, draw=darkred, fill=darkred!50, rounded corners, minimum width=0.5cm, minimum height=1cm, text centered]
        \tikzstyle{middle} = [rectangle, draw=gray, fill=gray!50, rounded corners, minimum width=0.35cm, minimum height=0.7cm, text centered]
        \tikzstyle{link} = [draw=darkgray, thick]

        % Nodes
        \node[endnode] (A) at (0, 0) {};      % Start node (bottom-left)
        \node[middle] (N1) at (1.5, 0.5) {};       % Middle node 1
        \node[middle] (N2) at (3, 1) {};     % Middle node 2
        \node[middle] (N3) at (4.5, 1.5) {};       % Middle node 3
        \node[middle] (N4) at (6, 2) {};     % Middle node 4
        \node[middle] (N5) at (7.5, 2.5) {};       % Middle node 5
        \node[middle] (N6) at (9, 3) {};     % Middle node 6
        \node[middle] (N7) at (10.5, 3.5) {};       % Middle node 7
        \node[endnode] (B) at (12, 4) {};       % End node (top-right)

        % Place text below each node
        \node at (A.south) [below=0.3cm] {Frankfurt};  % Text below City A
        \node at (N1.south) [below=0.3cm] {Schüchtern};  % Text below Node 1
        \node at (N2.south) [below=0.3cm] {Eiterfeld};  % Text below Node 2
        \node at (N3.south) [below=0.3cm] {Waltershausen};  % Text below Node 3
        \node at (N4.south) [below=0.3cm] {Erfurt};  % Text below Node 4
        \node at (N5.south) [below=0.3cm] {Eulau};  % Text below Node 5
        \node at (N6.south) [below=0.3cm] {Köckern};  % Text below Node 6
        \node at (N7.south) [below=0.3cm] {Schäpe};  % Text below Node 7
        \node at (B.south) [below=0.3cm] {Berlin};  % Text below City B

        % Curved Links using out=in syntax
        \draw[link] (A) to[out=30,in=180] (N1); % Curved path A to N1
        \draw[link] (N1) to[out=30,in=180] (N2); % Curved path N1 to N2
        \draw[link] (N2) to[out=30,in=180] (N3); % Curved path N2 to N3
        \draw[link] (N3) to[out=30,in=180] (N4); % Curved path N3 to N4
        \draw[link] (N4) to[out=30,in=180] (N5); % Curved path N4 to N5
        \draw[link] (N5) to[out=30,in=180] (N6); % Curved path N5 to N6
        \draw[link] (N6) to[out=30,in=180] (N7); % Curved path N6 to N7
        \draw[link] (N7) to[out=30,in=180] (B); % Curved path N7 to B

        % Add text above the links
        \node at (0.75, 0.75) {95.5 km}; % Text above link from Node 1 to Node 2
        \node at (2.25, 1.25) {67.5 km}; % Text above link from Node 2 to Node 3
        \node at (3.75, 1.75) {89 km}; % Text above link from Node 3 to Node 4
        \node at (5.25, 2.25) {46 km}; % Text above link from Node 4 to Node 5
        \node at (6.75, 2.75) {103.6 km}; % Text above link from Node 5 to Node 6
        \node at (8.25, 3.25) {81.6 km}; % Text above link from Node 6 to Node 7
        \node at (9.75, 3.75) {91.2 km}; % Text above link from Node 7 to City B
        \node at (11.25, 4.25) {90 km}; % Text above link from Node 7 to City B
        % % Links
        % \foreach \i/\j in {A/N1, N1/N2, N2/N3, N3/N4, N4/N5, N5/N6, N6/N7, N7/B}
        %     \draw[link] (\i) -- (\j);
        % Curved Links between nodes
      
    \end{tikzpicture}
    \caption{Linear network connecting the two German metropoles Frankfurt to Berlin with seven intermediate nodes. The links all slightly vary in distance and loss, and three different types of detectors are used in the entire network.}
    \label{fig:network_topology}
    \end{figure*}
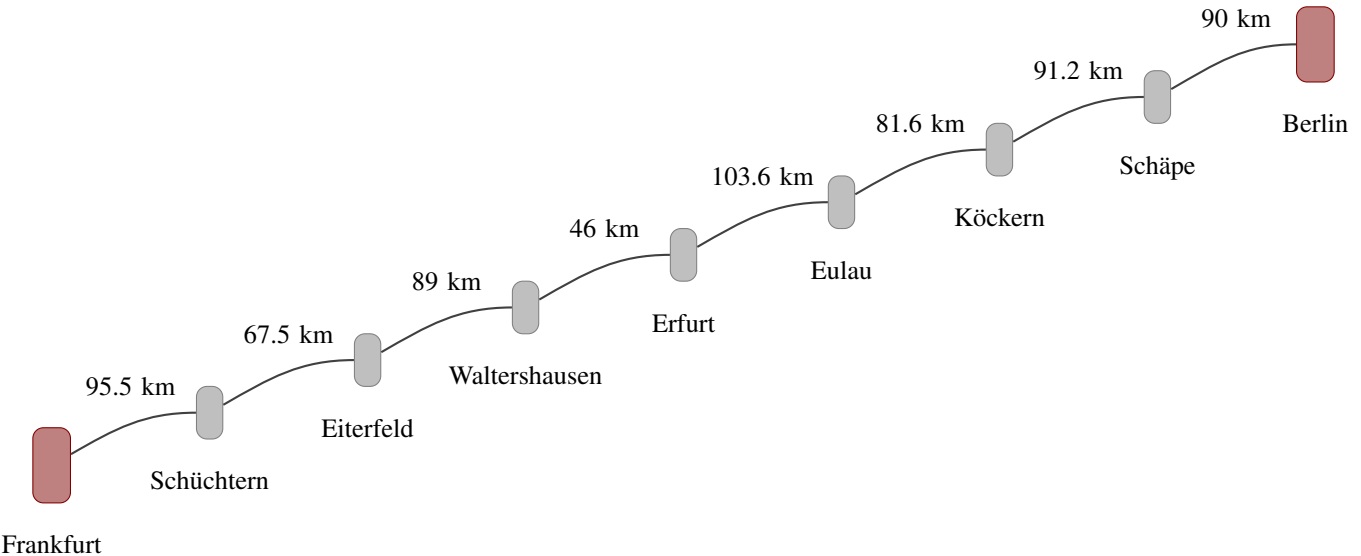

One such example is the long-distance quantum link between the two German metropoles—Berlin and Frankfurt—developed as part of the project Q-net-Q \footnote{funded by the EU and the German Federal Ministry for Education and Research.}. The network consists of seven intermediate trusted relay nodes, enabling key distribution over large distances (see Figure~\ref{fig:network_topology}). However, the available resources are limited to bipartite entanglement. This, along with the fact that the links are highly lossy, poses significant challenges for the efficient extraction of multi-partite entangled states.

In this work, we explore strategies to overcome these limitations by investigating how multipartite entangled states can be established in a network with realistic parameters. Given the significant loss in network links, we limit our initial focus to the generation of GHZ states among three network nodes. These states can then serve as a resource for more complex multi-party communication protocols, extending the network’s original goal of bipartite key exchange. Additionally, we explore the advantages and disadvantages of adding quantum memories (QMs) \cite{Heshami_2016, Lei:23} to this process and derive the necessary requirements for successfully generating high-fidelity entangled states in this linear quantum network.

Beyond state generation, we also explore the practical feasibility of cryptographic primitives that rely on multipartite entanglement. In particular, we assess the performance of Conference Key Agreement (CKA)\cite{Grasselli_2018, Epping_2017, Murta_2020}, anonymous Conference Key Agreement (ACKA)~\cite{Hahn:20, Grasselli:2022, Webb:24} and Quantum Secret Sharing (QSS)\cite{Hillery_1999, Karlsson_1999, memmen2023}, and point out when it is beneficial to use multi-partite entanglement. 

    \section{Network}
The quantum network considered in this work connects the two German metropoles Berlin and Frankfurt, via a 664.4 km long-distance fiber link consisting of seven intermediate trusted nodes (see Figure~\ref{fig:network_topology}). In its initial implementation, the network is designed to support Quantum Key Distribution (QKD) between all neighboring stations, ultimately enabling secure key exchange over the full end-to-end distance.
The individual links vary in length and loss, originating mostly from fiber attenuation, but also from the number of fiber connectors and splices used. Three types of detectors, each with different detection efficiency and dark count probability, are placed at the respective nodes. 
Furthermore, as part of the project, entangled photon-pair sources operating at a frequency of
$f = 40\cdot 10^{6}\text{ (pairs}/\text{sec)}$ are available and can be readily placed at the stations to generate the necessary bipartite entanglement. Quantum memories and two-qubit gates, such as entangling $C^Z$ gates, are not available within the scope of this project. For these components, we either use benchmark parameters from the literature or derive performance requirements needed to achieve the desired objectives.

\section{Establishing a 3-party GHZ state}
\label{Extraction_process}
\begin{figure}[h]
    \centering

    % First step (subfigure a)
    \begin{subfigure}[b]{\columnwidth}
        \centering
        \begin{tikzpicture}[scale=0.8]
            % Define the style for the nodes (qubits) with grey fill
            \tikzstyle{qubit}=[circle, draw, fill=gray!50, minimum size=5mm]
            
            % Place the nodes A, B, and C
            \node at (0, 0) {A};  % Node A
            \node[qubit] (qA) at (0, 1) {0}; % Qubit at A
            
            \node at (4.5, 0) {B};  % Node B
            \node[qubit] (qB1) at (4, 1) {1}; % Qubit 1 at B
            \node[qubit] (qB2) at (5, 1) {2}; % Qubit 2 at B
            
            \node at (9, 0) {C};  % Node C
            \node[qubit] (qC) at (9, 1) {3}; % Qubit at C
            
            % Draw edges to represent entanglement (CZ gates)
            \draw (qA) -- (qB1); % Edge between A and B (qA to qB1)
            \draw (qB2) -- (qC); % Edge between B and C (qB2 to qC)
        \end{tikzpicture}
        \subcaption{}  % Subcaption for the first step
    \end{subfigure}

    \vspace{1cm}  % Add vertical space between figures

    % Second step (subfigure b)
    \begin{subfigure}[b]{\columnwidth}
        \centering
        \begin{tikzpicture}[scale=0.8]
            % Define the style for the nodes (qubits) with grey fill
            \tikzstyle{qubit}=[circle, draw, fill=gray!50, minimum size=5mm]
            
            % Place the nodes A, B, and C
            \node at (0, 0) {A};  % Node A
            \node[qubit] (qA) at (0, 1) {0}; % Qubit at A
            
            \node at (4.5, 0) {B};  % Node B
            \node[qubit] (qB1) at (4, 1) {1}; % Qubit 1 at B
            \node[qubit] (qB2) at (5, 1) {2}; % Qubit 2 at B
            
            \node at (9, 0) {C};  % Node C
            \node[qubit] (qC) at (9, 1) {3}; % Qubit at C
            
            % Draw edges to represent entanglement (CZ gates)
            \draw (qA) -- (qB1); % Edge between A and B (qA to qB1)
            \draw (qB2) -- (qC); % Edge between B and C (qB2 to qC)
            \draw (qB1) -- (qB2); % Edge within node B
            
            % Labels for operations
            \node at (4.5, 2.25) {$C_Z$};
            \node at (5, 0.25) {$Y$};
        \end{tikzpicture}
        \subcaption{}  % Subcaption for the second step
    \end{subfigure}

    \vspace{1cm}  % Add vertical space between figures

    % Third step (subfigure c)
    \begin{subfigure}[b]{\columnwidth}
        \centering
        \begin{tikzpicture}[scale=0.8]
            % Define the style for the nodes (qubits) with grey fill
            \tikzstyle{qubit}=[circle, draw, fill=gray!50, minimum size=5mm]
            
            % Place the nodes A, B, and C
            \node at (0, 0) {A};  % Node A
            \node[qubit] (qA) at (0, 1) {0}; % Qubit at A
            
            \node at (4.5, 0) {B};  % Node B
            \node[qubit] (qB1) at (4, 1) {1}; % Qubit 1 at B
            
            \node at (9, 0) {C};  % Node C
            \node[qubit] (qC) at (9, 1) {3}; % Qubit at C
            
            % Draw edges to represent entanglement (CZ gates)
            \draw (qA) -- (qB1); % Edge between A and B (qA to qB1)
            \draw (qB1) -- (qC); % Edge between B and C (qB2 to qC)
        \end{tikzpicture}
        \subcaption{}  % Subcaption for the third step
    \end{subfigure}
    \caption{Steps to establish a three-party GHZ state starting from bipartite entanglement. In (a) the central node B establishes entanglement between itself and the outer nodes A and C respectively. As a second step (b), the central node connects the two local qubits $1$ and $2$ by applying a $C_Z$ gate on them and then measuring qubit $2$ in the $Y$ basis. This results in a three party linear cluster state as seen in (c), which is locally-equivalent to a GHZ state.}
    \label{fig:three_steps}
\end{figure}
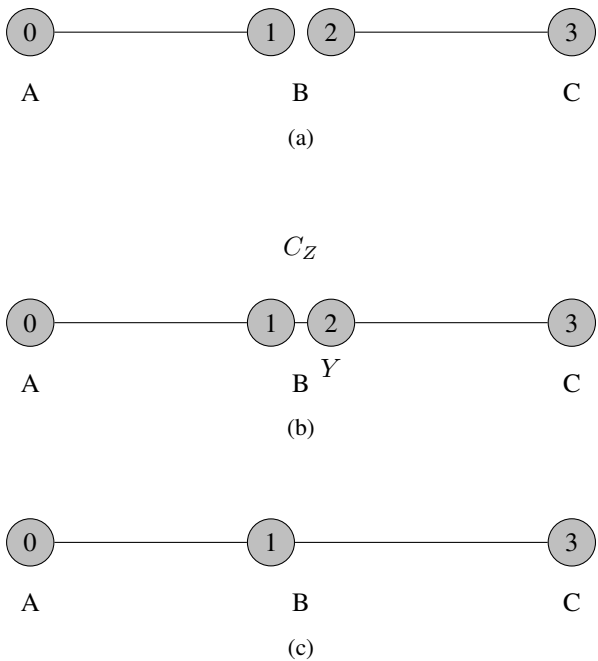

In this work, we will focus on the extraction of Greenberger-Horne-Zeilinger (GHZ) states \cite{GHZ89}, which are commonly used in many quantum cryptographic applications. An $n$-partite GHZ state reads:
\begin{align}
    \ket{\mathrm{GHZ}} = \frac{1}{\sqrt{2}} \left(\ket{0\ldots0} + \ket{1\ldots1}\right)_{1, \ldots, n}.
    \label{GHZ_state}
\end{align}
To establish a 3-party GHZ-state along a line, we begin with three nodes labeled A, B, and C. As a first step, the central node (B) establishes entanglement with its neighbors—the left node A and the right node C (see Figure~\ref{fig:three_steps}a).
Two entangled pairs are produced at station B in the state $\frac{1}{\sqrt{2}}(\ket{0+} + \ket{1-})$ and shared with nodes A and C\footnote{Note that any other Bell pair would also yield a state that would be locally equivalent to the GHZ state.}. Node B entangles its two local qubits (labeled 1 and 2 in Figure~\ref{fig:three_steps}b) by applying another $C^Z$ gate and measures qubit 2 in the $Y$-basis, preserving entanglement between the remaining three qubits.

The post-measurement state, depicted in Figure~\ref{fig:three_steps}c, depends on the outcome of the measurement (i.e., $\ket{\pm \text{y}}:= \frac{1}{\sqrt{2}}(\ket{0} \pm i \ket{1})$) as follows:
\begin{align} 
\ket{\phi_{\pm}}_{0,1,3} = \frac{1}{2} ((1 + i)\ket{+,0,\pm \text{y}} + (1 - i)\ket{-,1,\mp\text{y}})_{0,1,3}. 
\label{eq:LE_GHZ_pm} \end{align} 
It is easy to see that this state is locally equivalent to a GHZ state up to basis transformations and is a +1 eigenstate of the following stabilizers: 
\begin{align} 
&X_0 Z_1 \mathbbm{1}_3, \quad \mp X_0 \mathbbm{1}_1 Y_3, \quad - Y_0 X_1 Z_3, \quad \mp Y_0 Y_1 X_3,\notag \\ \mp &Z_0 X_1 X_3, \quad Z_0 Y_1 Z_3, \quad \mp \mathbbm{1}_0 Z_1 Y_3, \quad \mathbbm{1}_0 \mathbbm{1}_1 \mathbbm{1}_3. \label{stabilizers }\end{align}

\subsection{Storage of Quantum States}
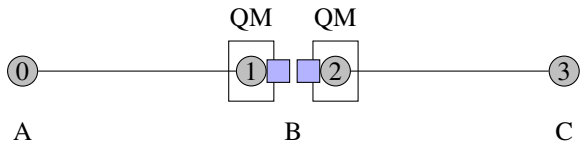
\begin{figure}[h]
    \centering
    \begin{tikzpicture}[scale=0.8]
        % Define styles
        \tikzstyle{qubit}=[circle, draw, fill=gray!50, minimum size=3mm, inner sep=1pt]
        \tikzstyle{memory}=[draw, rectangle, solid, minimum width=0.6cm, minimum height=0.8cm]
        \tikzstyle{source}=[draw, fill=blue!30, minimum width=0.3cm, minimum height=0.3cm]

        % Node A
        \node at (0, 0) {A};  
        \node[qubit] (qA) at (0, 1) {0};

        % Node B
        \node at (4.5, 0) {B};  
        \node[qubit] (qB1) at (3.8, 1) {1};
        \node[qubit] (qB2) at (5.2, 1) {2};

        % Node C
        \node at (9, 0) {C};  
        \node[qubit] (qC) at (9, 1) {3};

        % Edges
        \draw (qA) -- (qB1); 
        \draw (qB2) -- (qC); 

        % Quantum memories
        \node[memory, label=above:QM] at (3.8, 1) {};
        \node[memory, label=above:QM] at (5.2, 1) {};

        % Single shared source below qubits 1 and 2
        %\node[source, label=below:Source] at (4.5, 0.3) {};
        \node[source] at (4.25, 1) {};
        \node[source] at (4.75, 1) {};

    \end{tikzpicture}
    \caption{Adding quantum memories to the central station enables the storage of the local qubits $1$ and $2$. The central station creates bipartite entanglement and sends one half to the outer stations while keeping the other half locally. When these halves can be stored, the two links do not necessarily have to work simultaneously, which maximizes the per-round probability of creating the two required links. The blue squares symbolize the sources, which are assumed to be placed at the central node.}
    \label{fig:setup_QM}
    \vspace{1cm}
\end{figure}

If quantum memories are added to the network, this maximizes the probability of successful generation of the GHZ state shown in Figure~\ref{fig:three_steps}c. While we could equip all nodes with QMs, these are imperfect and decohere over time; therefore it is crucial to minimize their usage.

In fact, the single-qubit measurements on nodes A and C commute and therefore the respective qubits can be measured upon arrival. Node B does not necessarily need to be equipped with quantum memories either; the entangling operation could be performed 
while qubits $0$ and $3$ are still in transit to the end nodes. However, as both links are long-distance and highly lossy, the probability that they are both detected at nodes A and C is relatively low. Adding quantum memories at the central node maximizes the per-round success probability and thereby the state generation rate. The setup is illustrated in Figure~\ref{fig:setup_QM}. This configuration essentially corresponds to a quantum repeater setup~\cite{Luong_2016}, albeit with slightly different measurements.

\subsection{The Protocols}
Below we give two protocols, without and with memories, that establish a GHZ state between nodes A, B and C, when a source of bipartite entanglement is placed at node B. Since there is no way to know that a state has been created other than measuring in some basis, we assume that all parties perform a measurement that is compliant with the communication protocol/application for which the GHZ state is needed. This basis choice is for now irrelevant. The different applications will be discussed in Section~\ref{application_section}. 

\vspace{0.5em}
\begin{Protocol}
\textbf{Establishing a 3-party GHZ state without memories}
\begin{itemize}
    \item[\textit{1.}] \textit{Node B creates many bipartite states and shares them with nodes A and C. The qubits that are sent out are labeled $0$ and $3$ (for nodes A and C respectively), while $1$ and $2$ are kept at node B.}
    
    \item[\textit{2.}] \textit{While qubits $0$ and $3$ are in transit, node B applies a $C^Z_{(1,2)}$ gate to its local qubits $1$ and $2$. Qubit $2$ is measured in the $Y$-basis and $1$ is measured in a basis suitable for the specific application.}

    \item[\textit{3.}] \textit{Node A and C measure their qubits in a basis suitable for the subsequent application.}
    
    \item[\textit{4.}] \textit{After transmission of all the quantum states, the nodes exchange information about which qubits have been detected at nodes A and C, so that only coinciding detections are kept. Node B communicates for which rounds node C has to apply a classical bit flip to its data.}

\end{itemize}
\label{Extraction_protocol_memoryless}
\end{Protocol}

\vspace{0.1in}
\begin{Protocol}
\textbf{Establishing a 3-party GHZ state with memories}
\begin{itemize}
    \item[\textit{1.}] \textit{Node B creates many bipartite states and shares them with nodes A and C. The qubits that are sent out are labeled $0$ and $3$ (for nodes A and C respectively), while $1$ and $2$ are kept at node B.}
    
    \item[\textit{2.}] \textit{Qubits $1$ and $2$ are stored in memories at node B. If node A and C successfully measure a qubit, they inform node B who then applies a $C^Z_{(1,2)}$ gate on the respective halves of the detected states.}
    
    \item[\textit{3.}] \textit{Qubit $2$ is measured in the $Y$-basis, and the outcome is stored. Qubit $1$ is measured in whatever basis is needed for the specific application. }
    
    \item[\textit{4.}] \textit{Node B communicates for which rounds node C has to apply a classical bit flip to its data.}
\end{itemize}
\label{Extraction_protocol_memory}
\end{Protocol}

\section{Models}
\label{model_section}
In this section, the models used for the different components are described. We outline which parameters are taken from the actual experimental implementation of the testbed, which are taken from literature benchmarks, and which are treated as variable parameters in our performance analysis.
\paragraph*{Loss and dark counts}
The photons that are transmitted through the optical fiber experience exponential loss; we denote the optical loss of the fiber due to attenuation, connectors and splices with $p_{\mathrm{LINK}_{AB}}$ and $p_{\mathrm{LINK}_{BC}}$. Furthermore, all detection setups have a certain efficiency, which we denote as $\eta_{A}, \eta_{B}$, and $\eta_{C}$. The optical fiber loss and detection efficiencies are obtained from experimental measurements conducted at the testbed. In the memory-less case, we can define the probability of a photon being detected at setup $A$, $B$ and $C$ as:
\begin{align*}
\xi_{A} &= \eta_{A} \cdot p_{\mathrm{LINK}_{AB}} \notag \\
\xi_{B} &= \eta_{B} \notag \\
\xi_{C} &= \eta_{C} \cdot p_{\mathrm{LINK}_{BC}}.
\label{xi}
\end{align*}
Note that for node $B$, this only includes the local detection efficiency, while for the other two, this also includes the link transmission. When quantum memories are used, the expressions for $\xi_{A}$ and $\xi_{C}$ remain unchanged, while for the central station, the memory efficiency $\eta_{\mathrm{QM}}$ must be included:
\begin{align*}
\xi_{B, \mathrm{QM}} = \eta_{B} \cdot \eta_{\mathrm{QM}}.
\end{align*}
In this work, we assume $\eta_{\mathrm{mem}}=0.9$, an optimistic but very realistic goal for near-term quantum memories \cite{Schupp:2021}.
We can then define $\xi'_{j}$ to be the probability that the detector at node $j\in\{A,B,C\}$ clicks:
\begin{equation}
    \xi'_{j} =1 - (1 - \xi_{j})(1 - p_{d_j})^2,
    \label{xi_prime}
\end{equation}
where $p_{d_j}$ is the probability that detector $j$ records a detection that is due to a dark count and not an actual photon. $p_{d_j}$ is specific to the type of detector installed and operational at the respective station. Note here that in our modeling, dark counts are the only source of noise that stems from the measurement and are modeled as single-qubit depolarization:
\begin{align}
\mathcal{E} (\ket{\psi}\bra{\psi}) = (1 - \alpha) \ket{\psi}\bra{\psi} + \alpha \frac{\mathbbm{1}}{2}.
\label{depol_map}
\end{align}
In the case of dark counts, the depolarization parameter $\alpha$ for each node $j\in\{A,B,C\}$ is \cite{Luong_2016}:
\begin{align}
\alpha_{\mathrm{DC}_{j}} = 1 - \frac{\xi_{j} (1 - p_{d_j})}{\xi'_{j}}.
\end{align}
Dark counts are more prevalent at the outer nodes, as the overall transmission is lower due to the losses in the channel.

\paragraph*{Depolarization on fiber}
In the context of optical fibers, every qubit that enters the fiber is also subject to single-qubit depolarization (Equation~\eqref{depol_map}), where $\alpha=f_{\mathrm{D}}$. This parameter is not predetermined and will vary during our performance analysis.
\paragraph*{Imperfect gates}
Imperfections of gates are also modeled as depolarization\cite{Epping_2017}. When a gate fails, which happens with probability $f_{\mathrm{G}}$, both the control and target qubit (qubit $i$ and $j$) are traced out and replaced by the maximally mixed state on their respective subsystems
\begin{align}
    C^Z_{i,j} (\rho) = (1 - f_{\mathrm{G}}) \, C^Z_{i,j} \, \rho \,C^Z_{i,j} + \frac{f_{\mathrm{G}}}{4} \text{Tr}_{i, j} \big(\rho\big) \otimes \mathbbm{1}_{i, 2j}.
\end{align}
Again, this parameter is not predetermined and will vary during our performance analysis.
\paragraph*{Memory decoherence}
Memory decoherence is modeled as a time-dependent dephasing channel on the $i$-qubit of the initially stored state $\rho$\cite{Razavi:2009}
\begin{align}
\Gamma_i (\rho) = (1 - \lambda_{\text{dp}}(t))\rho + \lambda_{\text{dp}}(t) Z_i \rho Z_i,
\end{align}
where 
\begin{align}
\lambda_{\text{dp}}(t) = \frac{1 - e^{-t/T_2}}{2}.
\end{align}
Here, $t$ is the time that the respective qubit is stored in the QM and $T_2$ is the internal dephasing time, an indicator for the memory quality. We will consider a dephasing time of $T_2 = 2.5$s (as reported in \cite{Olmschenk:2007} for a trapped-ion qubit).
\vspace{0.1in}

In order to determine the amount of noise the photons experience while being stored in the QM, we need to evaluate how long they are stored on average. In our setting, the sources are placed at the central station. One trial then consists of a bipartite entangled state being created and sent out to the respective stations. Additionally, the central station has to receive confirmation that the outer stations have successfully detected their photon in order to determine whether the local photon needs to be stored. This yields 
\begin{align} \tau_{A} &= T_p + \frac{2 L_{A}}{c} \notag \\ 
\tau_{C} &= T_p + \frac{2 L_{C}}{c}, 
\end{align} where $T_p$ is the preparation time for a bipartite entangled state and is related to the frequency $f$ via $T_p=\frac{1}{f}$, and $L_{A}$ and $L_{C}$ are the distances between node $B$ and nodes $A$ and $C$ respectively. The speed of light in optical fiber is $c = 2\cdot10^{5} \frac{\mathrm{km}}{\mathrm{s}}$. On average, the station that is farther from the central station will detect a photon later than the one that is closer. Let us assume that this is station C. Then, the photon from the same Bell pair as the photon at station C will experience dephasing for 
\begin{align} t_C &= \frac{2 L_{C}}{c}. 
\end{align} 
We consider simultaneous loading, meaning that the central station tries to establish entanglement with both outer nodes at the same time. The photon that is part of the same Bell pair as the photon at station A then dephases for 
\begin{align} 
t_A &= \lvert N_A - N_C \rvert \tau_C + \frac{2 L_{A}}{c}, 
\end{align} 
where $N_A$ and $N_C$ denote the number of attempts needed for the detectors at stations A and C to click once, respectively. They are random variables with success probabilities $\xi'_A$ and $\xi'_C$, respectively.
The expectation value of this was evaluated in \cite{Panayi_2014} to be
\begin{align}
    \mathbb{E}\left(e^{-t_{A}/T_2}\right) =
    &\frac{\xi'_{C} \exp\left(-\frac{2L_{A}}{c}\right)}
    {\xi'_{A} + \xi'_{C} - \xi'_{A} \xi'_{C}} 
    \biggl[
        \frac{1}{1 - e^{-\tau_{C}/T_2}(1 - \xi'_{A})}
        \\
        &+ \frac{1}{1 - e^{-\tau_{C}/T_2}(1 - \xi'_{C})}
        - 1 
    \biggr]. \notag
\end{align}

\section{Performance}
While the network in Figure~\ref{fig:network_topology} was originally designed for bipartite key exchange between Berlin and Frankfurt, here we explore which other, more complex communication tasks can be performed on smaller segments of the network. We focus on segments consisting of three nodes: Berlin–Sch{\"a}pe–K{\"o}ckern, K{\"o}ckern–Eulau–Erfurt, Erfurt–Waltershausen–Eiterfeld, and Eiterfeld–Sch{\"u}chtern–Frankfurt. The middle node always corresponds to the central station B, while the other two represent the outer nodes A and C. When evaluating the performance of our protocol, we focus on two key indicators: the probability of successfully generating a 3-party GHZ state and its quality. Ideally, both of these measures should be high. However, as we will see, there is a trade-off between achieving a high generation rate and maintaining high-quality entangled states. We will characterize the quality of the generated states by their fidelity with respect to the ideal target state.
\subsection{Generation rate}
\label{probability_section}

\subsubsection{Memoryless case}
Without the addition of QMs, all four detectors must register a click during the same round. This corresponds to a yield:
\begin{align}
Y = \xi'_{A} (\xi'_B)^2 \xi'_{C}. 
\label{yield_memoryless}
\end{align}
As the links in our network are long-distance links, this probability will be very small. 
\subsubsection{Using Quantum Memories}

If quantum memories are used, qubits can be stored at the central node when the two links are not established simultaneously. This increases the overall probability of a successful round. $N_A$ and $N_C$ denote the random variables representing the number of attempts required such that the outer nodes A and C successfully detect a click, due to dark counts or proper detections. Hence, the corresponding success probabilities are again $\xi'_A$ and $\xi'_C$ (see Equation~\eqref{xi_prime}). The expected number of attempts in the memory-assisted case was evaluated in \cite{Panayi_2014}. Additionally, the measurements at the central node also need to be successful (which each happens with success probability $\xi'_{B, \mathrm{QM}}$). The yield when using QMs is then:
\begin{eqnarray}
Y_{\mathrm{QM}} &=& \frac{(\xi'_{B, \mathrm{QM}})^2}{\mathbb{E}(\max(N_A, N_C))} \\
&=& (\xi'_{B, \mathrm{QM}})^2 \bigg( \frac{1}{\xi'_{A}} + \frac{1}{\xi'_{C}} - \frac{1}{\xi'_{A} + \xi'_{C} - \xi'_{A} \xi'_{C}} \bigg)^{-1}. \notag
\label{yield_memory}
\end{eqnarray}

In Table~\ref{tab:probabilities_sublinks}, the probabilities for the generation of a three-partite entangled state are displayed for the four different segments consisting of three nodes, with and without the use of QMs. As the links are all in the high-loss regime, the probability of successful generation without the use of QMs is very small. Adding QMs can boost the probability by up to three orders of magnitude, depending on the specific links and their parameters. 
\begin{table}[h]
    \centering
    \resizebox{\columnwidth}{!}{%
        \begin{tabular}{|c|c|c|c|}
            \hline
        Link & $Y$ & $Y_{\mathrm{QM}}$ & $Y_{\mathrm{QM}}/Y$\\
        \hline
        Berlin–Sch{\"a}pe–K{\"o}ckern & $3.6\cdot 10^{-7}$ & $1.3\cdot 10^{-4}$ & $359$ \\
        K{\"o}ckern–Eulau–Erfurt & $9.3\cdot 10^{-9}$ & $5.1\cdot 10^{-6}$ & $543$ \\
        Erfurt–Waltershausen–Eiterfeld & $1.9\cdot 10^{-7}$ & $4.1\cdot 10^{-6}$ & $21$ \\
        Eiterfeld–Sch{\"u}chtern–Frankfurt & $6.2\cdot 10^{-9}$ & $2.7\cdot 10^{-6}$ & $441$ \\
        \hline
        \end{tabular}
    }
    \caption{Table listing the yields for the three node configurations with and without memories, and the improvement due to the usage of QMs.}
    \label{tab:probabilities_sublinks}
\end{table}

\subsection{Fidelities}
The fidelity of the distributed states serves as a key indicator for their quality. In our scenario, several sources of noise influence the final fidelity (see Section \ref{model_section}). These are the depolarization due to optical fiber, dark counts in the detectors for all required measurements and imperfections of the $C_Z$-gates. Additionally, when using QMs, we also need to take into account the introduced dephasing. 

\begin{figure*}[t]
    \centering

    \begin{subfigure}{0.9\textwidth}
        \centering
        \includegraphics[width=\linewidth]{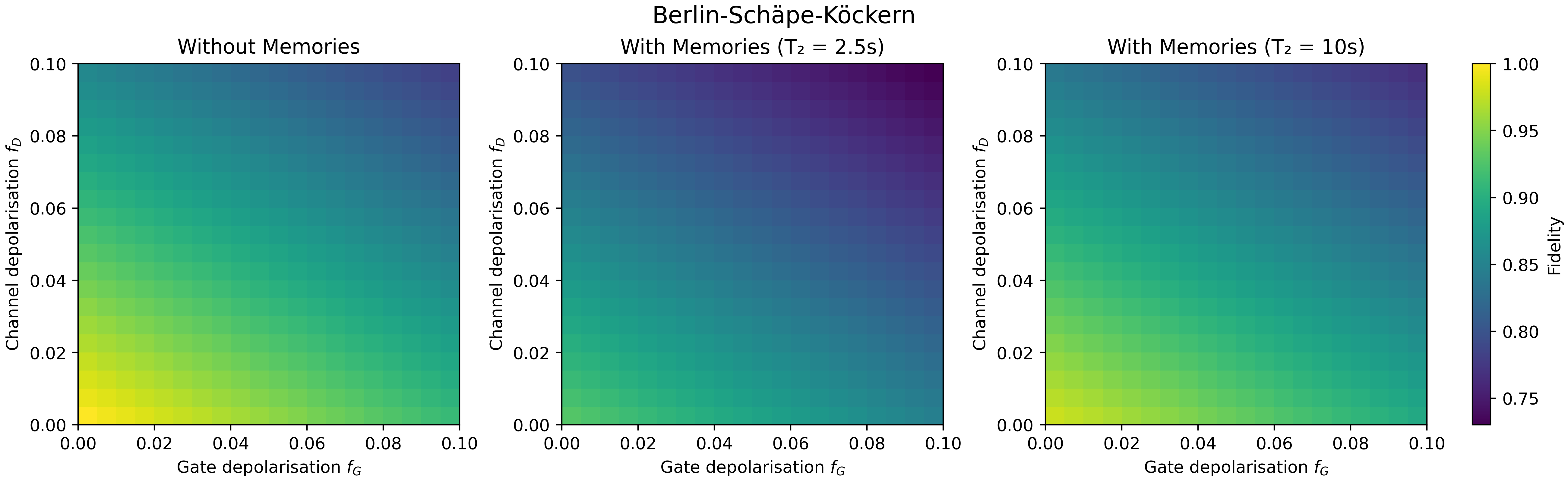}
        %\caption{Caption for subfigure (a).}
        \label{fig:fig1a}
    \end{subfigure}
    
    \vspace{1em}

    \begin{subfigure}{0.9\textwidth}
        \centering
        \includegraphics[width=\linewidth]{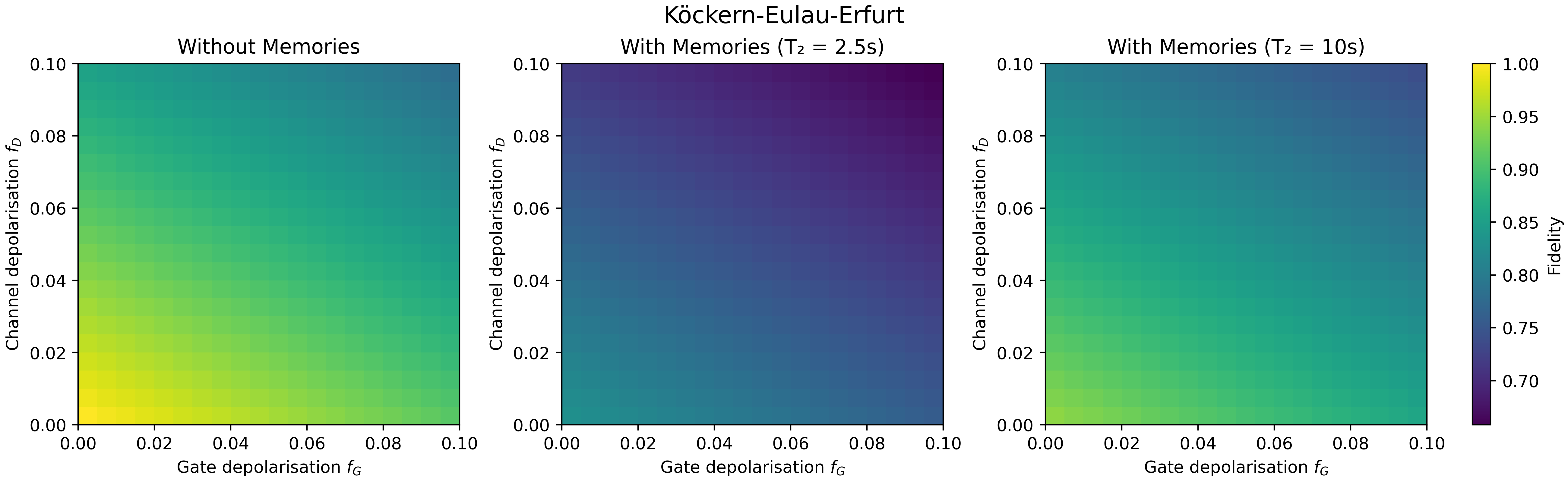}
        %\caption{Caption for subfigure (b).}
        \label{fig:fig1b}
    \end{subfigure}
    
    \vspace{1em}

    \begin{subfigure}{0.9\textwidth}
        \centering
        \includegraphics[width=\linewidth]{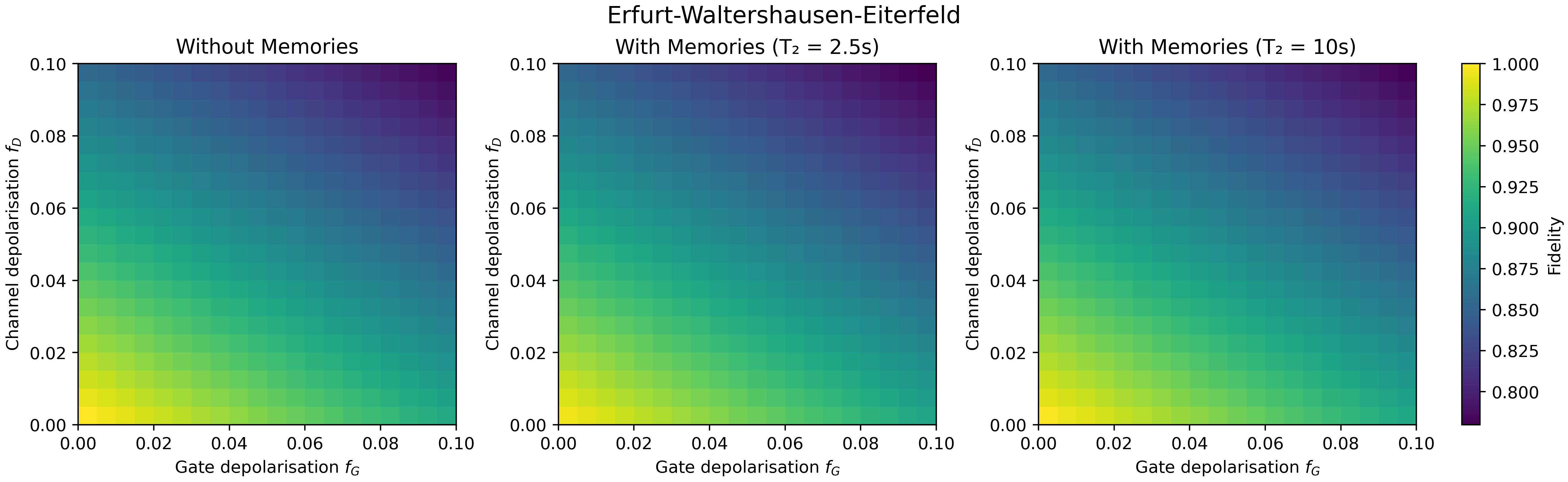}
        %\caption{Caption for subfigure (c).}
        \label{fig:fig1c}
    \end{subfigure}
    
    \vspace{1em}

    \begin{subfigure}{0.9\textwidth}
        \centering
        \includegraphics[width=\linewidth]{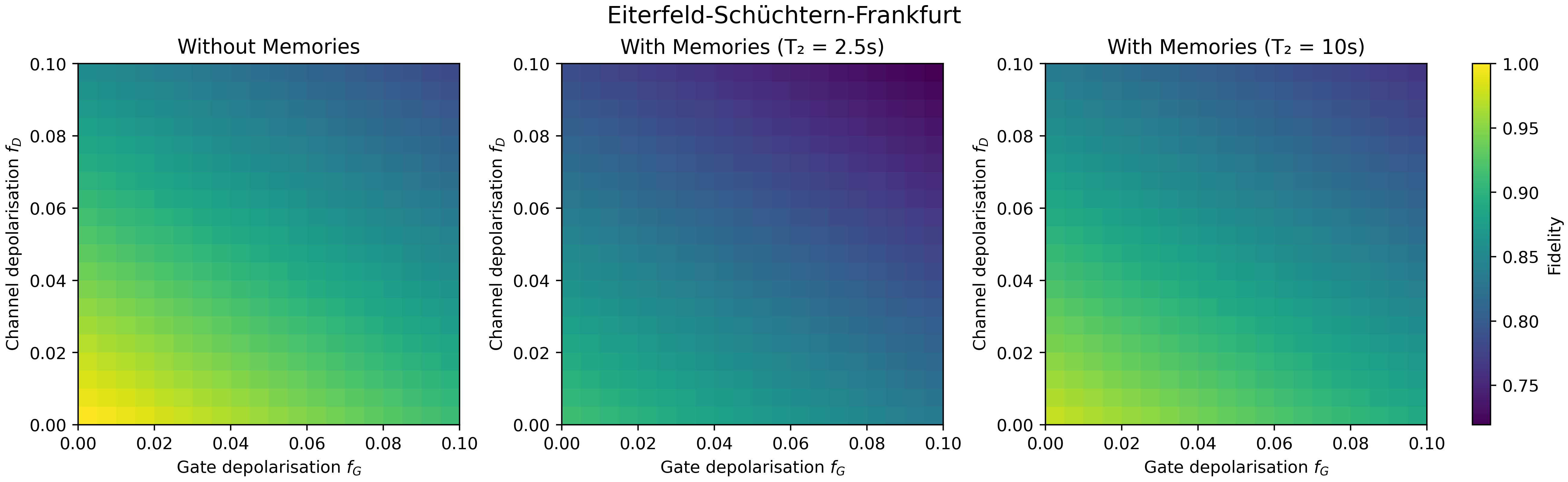}
        %\caption{Caption for subfigure (d).}
        \label{fig:fig1d}
    \end{subfigure}

    \caption{Fidelities of the established 3-partite GHZ state in dependence of the gate failure probability $f_{\mathrm{G}}$ and the channel depolarization $f_{\mathrm{D}}$ for the different 3-party configurations.}
    \label{fig:fidelities_4}
\end{figure*}

All the above operations on a shared state are modeled using the quantum network simulator requsim\cite{Wallnofer_2024}. Then, the fidelity of the noisy state $\rho$ with the perfect state\footnote{Note here that without loss of generality, we consider only the case of a +1 measurement on node 2, as a -1 measurement can also be taken into account by a simple bit flip on the outcome.} in Equation~(\ref{eq:LE_GHZ_pm}) is calculated via:
\begin{align}
F = \bra{\phi_+} \rho \ket{\phi_+}.
\end{align}
The fidelities in dependence of the channel depolarization probability $f_{\mathrm{D}}$ and the 
$C_Z$-gate failure probability are shown in Figure~\ref{fig:fidelities_4}, for both settings. In the setting without QMs, the achievable fidelities are always maximal, as the use of QMs introduce more noise in the form of dephasing. However, this advantage comes at the expense of a significantly reduced generation rate (see Section~\ref{probability_section} for comparison). 
For the case including the use of QMs, we plot one graph assuming a realistic internal dephasing time of $T_2=2.5$s, and another assuming an optimistic $T_2=10$s. While the realistic $T_2$ in most cases results in noticeably lower fidelities compared to the case without QMs, the optimistic $T_2$ yields fidelities that are only marginally lower. An exception here is the link configuration Erfurt-Waltershausen-Eiterfeld, where a realistic memory quality already yields fidelities that are comparable to the case without QMs, yet with the advantage in terms of the generation rate. This is because that configuration consists of one link that is notably shorter, and hence has a significantly lower loss due to fiber attenuation (see Figure \ref{fig:network_topology}). Our results suggest that modest improvements in quantum memory technology could enable the generation of high-fidelity states within the existing network topology.

\section{Applications in Multipartite Quantum Communication}
\label{application_section}
Multipartite entangled states such as GHZ states are foundational for several quantum communication primitives. Two examples are \emph{Conference Key Agreement} (CKA) and \emph{Quantum Secret Sharing} (QSS).
While CKA is an extension of quantum key distribution (QKD) where a dealer distributes a shared secret key to multiple trusted participants, QSS allows a dealer to distribute a classical secret among untrusted participants, where only collaboration among all parties enables reconstruction of the secret. For both primitives, GHZ states can be used as a resource.
However, in a star-like network topology—where the dealer is centrally connected to all participants—these primitives can also be implemented using only bipartite entanglement. For instance, by establishing independent QKD links with each participant, the dealer can transmit either a shared conference key (for CKA) or separate secret shares (for QSS) via those pre-established links. This requires more network uses compared to using multi-partite entanglement, but is, in most cases, nevertheless the more practical solution as the generation of bi-partite entanglement is significantly less challenging experimentally.

An interesting paradigm where the benefits of multi-partite entanglement cannot be reproduced using bi-partite states is \emph{anonymous} CKA (ACKA)~\cite{Hahn:20, Grasselli:2022, Webb:24}. There, we have the additional requirement that the identities of the communicating parties should remain hidden. This is effectively implemented by a subroutine in which the subset of users aiming to communicate anonymously extracts a smaller GHZ state from an originally shared larger GHZ state. Although in principle we could also use bipartite entanglement\cite{Broadbent:2007, Huang:2022}, we would need to establish links between all pairs of participants to ensure anonymity, which would be extremely resource consuming.

In our setting, which is a three-node linear sub-network, the topology is effectively equivalent to a star-like configuration with the central node being the dealer. As we create multipartite entanglement from bipartite links, involving additional (and noisy) operations, there will not be a performance advantage when using multi-partite entanglement for CKA or QSS compared to directly using the bipartite links.

However, scaling to more than three parties, a linear and a star-like topology are no longer equivalent. Here, generating genuine multi-partite entanglement between the participants of a protocol, becomes essential. For instance, while CKA is feasible in a linear network when all nodes participate (as the key can be routed through each node of the network), when only a subset of the nodes wants to establish a key, genuine multipartite entanglement becomes necessary. Multipartite entangled states are also necessary for QSS, since the participants cannot be trusted to route information through the network. 

If we examine more closely the use of general GHZ states (see Equation~\eqref{GHZ_state}) for CKA, ACKA and QSS, we first note that in the $Z$ basis, the GHZ state exhibits individual correlations across all subsystems. This is why this basis can be used for key generation in CKA and ACKA. Conversely, a collective measurement in the $X$ basis is a stabilizer of the GHZ state and thus the parity of all outcomes is always positive. This type of correlation is needed to distribute the shares of the key for QSS. For parameter estimation, the roles of the bases are then reversed: in QSS, the $Z$ basis is used to estimate individual errors between the dealer and each participant, while for CKA and ACKA, the $X$ basis is used for parameter estimation.

In the asymptotic limit, the key rate for QSS \cite{memmen2023}, CKA \cite{Murta_2020}, and ACKA \cite{Grasselli:2022} is given by:
\begin{align}
    r_{\infty} = Y\left(1 - h(Q_X) - \max_i h(Q_{AB_i})\right),
\end{align}
where $Q_X$ the quantum bit error rate (QBER) for the collective $X$ measurement, and $Q_{AB_i}$ the individual QBER between the dealer and participant $i$ when measuring in the $Z$ basis.

In our case, slight modifications are required, as at the end of our protocol, we hold a state that is only locally equivalent to the GHZ state. The individual correlations can be accessed by measuring $X_0 Z_1 Y_3$. For the parity measurement, we consider one of the stabilizers (for instance $Z_0 Y_1 Z_3$) with support on all three subsystems (see Section~\ref{Extraction_process}).
The bipartite QBER can then be computed as:
\begin{align}
    \max_i Q_{AB_i} = 1 - \bra{\psi^+}\rho\ket{\psi^+} - \bra{\psi^-}\rho\ket{\psi^-},
\end{align}
where $\ket{\psi^{\pm}} = \frac{1}{2} ((1 - i)\ket{+,0,+\text{y}} \pm (1 + i)\ket{-,1,-\text{y}})$.

The QBER of the parity measurement, $Q_{X}$, is evaluated by summing the probabilities of all outcomes corresponding to a negative parity when measured in $Z_0 Y_1 Z_3$, since a positive parity is expected for the ideal state.

\begin{figure*}[t]
    \centering

    \begin{subfigure}{0.9\textwidth}
        \centering
        \includegraphics[width=\linewidth]{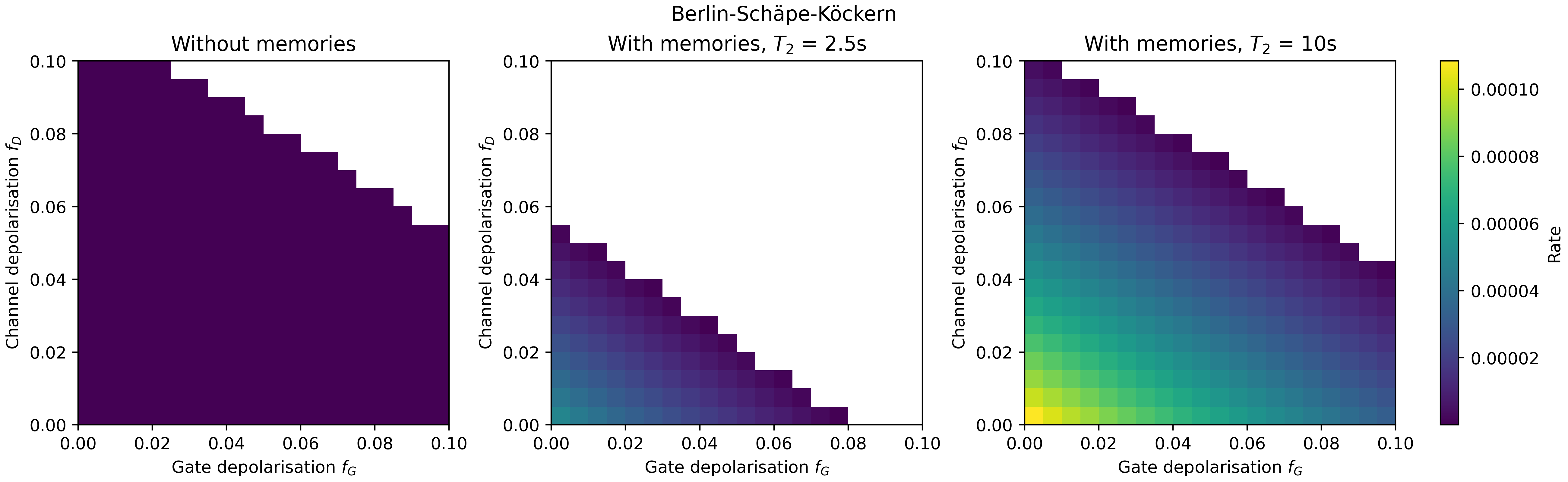}
        %\caption{Caption for subfigure (a).}
        \label{fig:fig2a}
    \end{subfigure}
    
    \vspace{1em}

    \begin{subfigure}{0.9\textwidth}
        \centering
        \includegraphics[width=\linewidth]{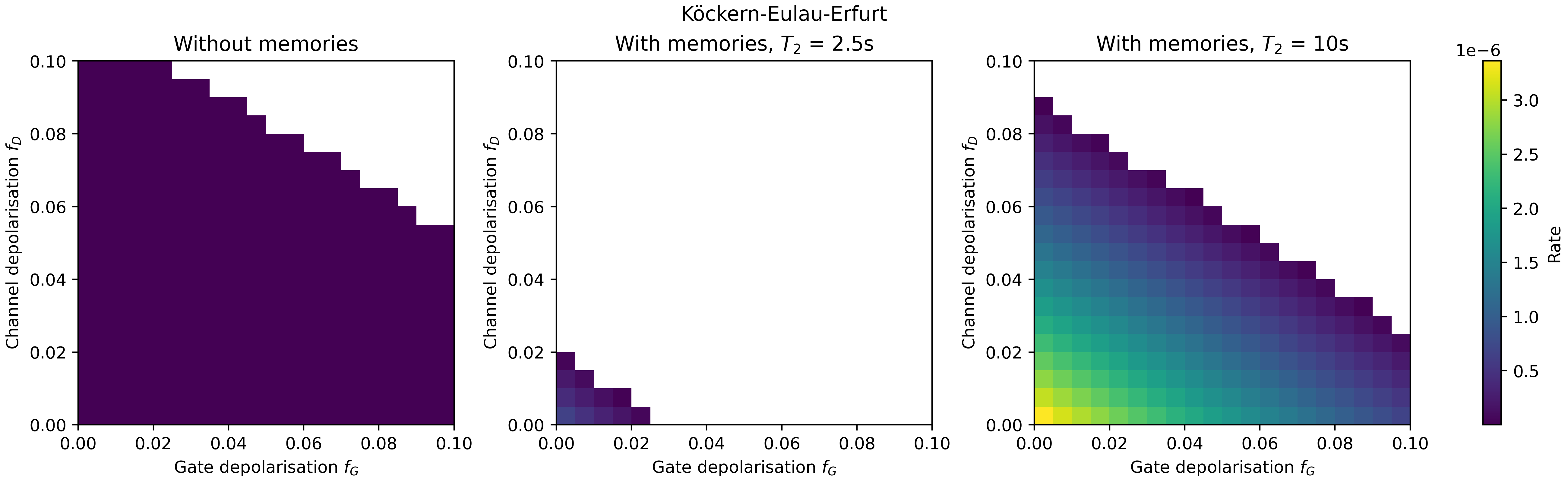}
        %\caption{Caption for subfigure (b).}
        \label{fig:fig2b}
    \end{subfigure}
    
    \vspace{1em}

    \begin{subfigure}{0.9\textwidth}
        \centering
        \includegraphics[width=\linewidth]{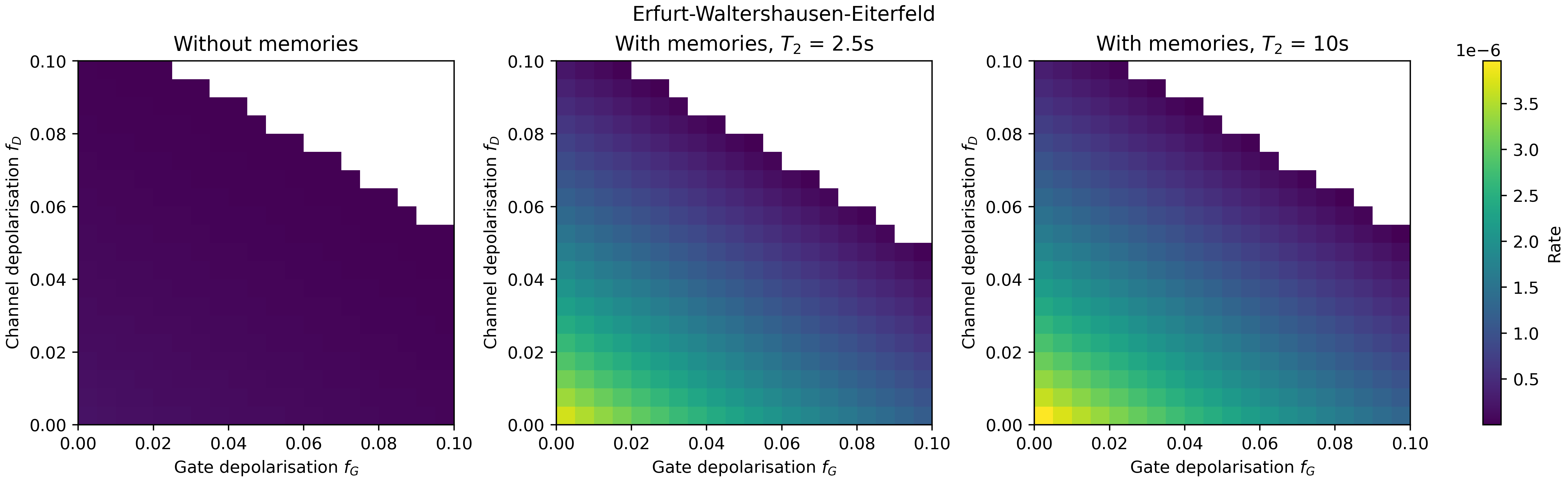}
        %\caption{Caption for subfigure (c).}
        \label{fig:fig2c}
    \end{subfigure}
    
    \vspace{1em}

    \begin{subfigure}{0.9\textwidth}
        \centering
        \includegraphics[width=\linewidth]{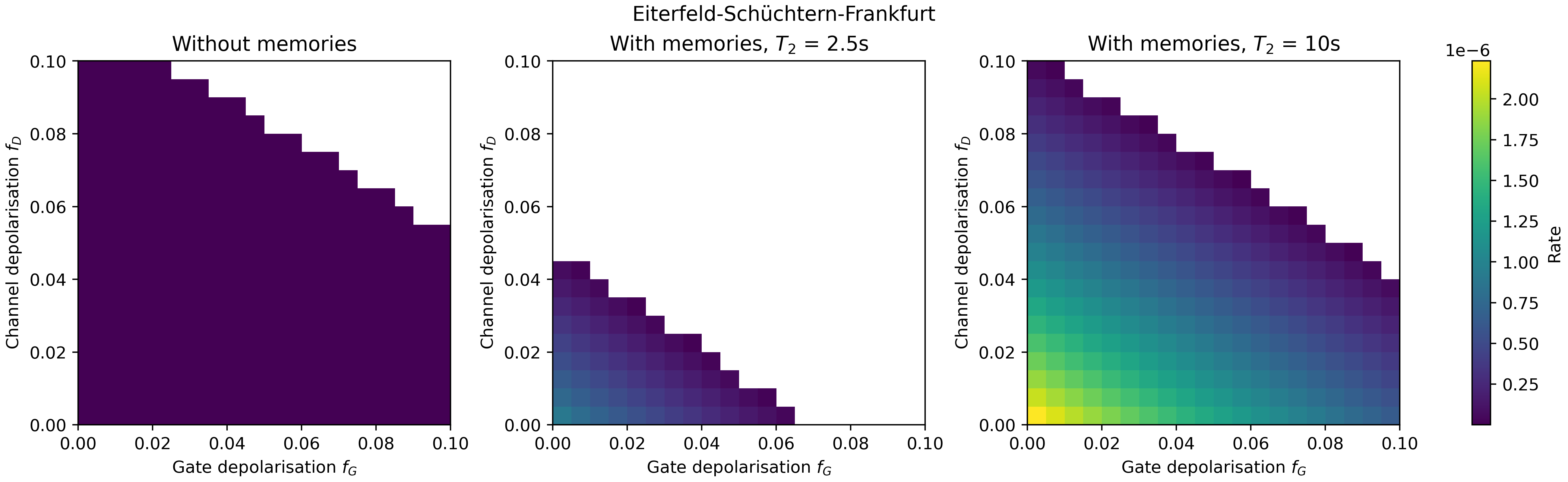}        %\caption{Caption for subfigure (d).}
        \label{fig:fig2d}
    \end{subfigure}

    \caption{Distillable rates for CKA, ACKA and QSS using the established 3-partite GHZ in dependence of the gate failure probability $f_{\mathrm{G}}$ and the channel depolarization $f_{\mathrm{D}}$ for the different 3-party configurations. For every sub-network, the left graph shows the case without QMs, while the middle and right one show the case using memories with different qualities ($T_2$). $T_2=2.5$s can be considered as a memory quality that can be achieved nowadays, while $T_2=10$s might be achievable with technical improvements.}
    \label{fig:CKA_rates_4}
\end{figure*}

Achievable key rates for CKA, ACKA, and QSS over the four subnetworks, each consisting of three network nodes, are depicted in Figure~\ref{fig:CKA_rates_4}, again for two quantum memory qualities $T_2$. If no data is shown, key distillation was not possible for those parameters.

Two competing effects influence the key rates. On the one hand, in the absence of quantum memories, the generation rates are extremely low due to the inherently low probability of successfully generating a three-partite entangled state. On the other hand, the fidelities in this setting are higher, because no additional noise is introduced by imperfect memories. This results in lower QBERs and, consequently, higher error tolerance with respect to gate errors $f_{\mathrm{G}}$ and channel depolarization $f_{\mathrm{D}}$.

In contrast, when QMs are used, the fidelities are in general lower due to the additional dephasing noise, leading to higher QBERs. Consequently, the parameter regimes in which key distillation remains possible becomes smaller. This effect is particularly visible for realistic memory coherence times $T_2 = 2.5$s, where key rates are only distillable in a very narrow range of the error parameters $f_{\mathrm{G}}$ and $f_{\mathrm{D}}$. However, assuming an optimistic memory quality $T_2 = 10$s almost recovers the thresholds in the case of no QMs, with a notable increase in the key rates. 

\section{Discussion}
In this work, we evaluated the extraction of a three-partite GHZ state in a real-world linear network. This involved analyzing the feasibility and efficiency of generating entanglement across three nodes in the network. We focused on two key performance indicators: the generation rate and the fidelity of the generated states, both with and without the use of quantum memories. While the inclusion of quantum memories significantly increased the generation rate—by nearly two orders of magnitude—it also introduced additional noise, reducing the fidelity of the resulting states. This reduction was significant for realistic memories, however, moderate technical improvement already yielded fidelities close to the case without QMs.

Furthermore, we examined the parameter regimes in which the generated resource states are suitable for implementing CKA, ACKA, and QSS. We evaluated the performance advantage of these multipartite approaches compared to schemes relying solely on bipartite entanglement and highlighted the conditions under which multipartite entanglement becomes necessary and yields better results.

Looking ahead, it would be interesting to scale up our approach to larger linear networks, where multipartite entanglement seems to offer an advantage compared to bipartite. Unfortunately, our current network is highly lossy, and three-partite GHZ states can only be established with tight constraints on the additional operations. Significant improvements in the experimental parameters would be required for practical implementations in such scenarios.

Finally, exploring the generation of other classes of entangled states and their potential applications is another promising direction~\cite{markham_graph_2008}. Future work could also investigate network variations, such as different placements of entanglement sources or the introduction of cut-off times for quantum memories to balance noise and efficiency. In addition, exploring alternative merging strategies~\cite{Wallnofer:2016} and more detailed models of quantum memory implementations, including photon-atom interface challenges, would be valuable next steps. Altogether, our results provide a compelling example of multipartite entanglement extraction in a real-world quantum network and point toward promising future applications of such technologies.
\clearpage
\section{Acknowledgements}
The authors acknowledge funding via the Q-net-Q Project (supported by the BMBF and EU’s Digital Europe Program No. 101091732), the BMBF project tubLANQ.0 (grant No. 16KISQ087K) and the Emmy Noether DFG (grant No. 418294583).

\bibliography{GS_QSS}
\bibliographystyle{ieeetr}
\end{document}